**Software paper for submission to the Journal of Open Research Software**

**(1) Overview**

**Title**
COMPLEX-IT: A Case-Based Modeling and Scenario Simulation Platform for Social Inquiry

**Paper Authors**
1. Schimpf, Corey;
2. Castellani, Brian

**Paper Author Roles and Affiliations**
1. Learning Analytics Scientist, The Concord Consortium, Boston USA.
2. Professor, Department of Sociology, Durham University, UK.

**Abstract**
COMPLEX-IT is a case-based, mixed-methods platform for social inquiry into complex data/systems, designed to increase non-expert access to the tools of computational social science (i.e., cluster analysis, artificial intelligence, data visualization, data forecasting, and scenario simulation). In particular, COMPLEX-IT aids social inquiry though a heavy emphasis on learning about the complex data/system under study, which it does by (a) identifying and forecasting major and minor clusters/trends; (b) visualizing their complex causality; and (c) simulating scenarios for potential interventions. COMPLEX-IT is accessible through the web or can be run locally and is powered by R and the Shiny web framework.

**Keywords**
Complex systems; computational modeling, case-based methods; mixed-methods research; social complexity theory; evaluation research, machine learning, data forecasting, data mining, data visualization.

Introduction

**I. The Challenges of studying complex data**

Many of the data sets and topics that social scientists, policy analysts, decision-makers and public employees regularly struggle to understand, whether they are ecological, health, economic or other types, may be best described as complex; that is, they are case-based (as opposed to variable focused), multi-dimensional, multi-level, dynamic, nonlinear, evolving along multiple trajectories (often in real-time), self-organizing, emergent, network-like in structure, and geospatially and contextually (path) dependent[1-2]. They also sit at the intersection (nexus) of several interconnected areas of study; and they often require the additional knowledge of how to influence, change or alter the course of these complex data (again, in real-time) – particularly in such areas as applied research, healthcare, education, public infrastructure, social services, and policy and program evaluation.

Addressing this level of data complexity presents a series of methodological challenges.  First, most users in the public sector and the social and health sciences are only trained in the conventional methods of statistics or qualitative inquiry. Second, even for those aware of the recent developments in computational modeling, data mining or big data analytics, and the complexity sciences – which are at the forefront of dealing with such complex data – these tools and technique remain beyond everyday usage. Which, in turn, creates a third challenge, as users wanting to employ these new techniques often become over-reliant upon specialists that have no background or expertise in the topics they are being asked to model. Fourth, even for those skilled in the latest developments in computational modeling etc., there is presently little software available that integrates these techniques into a dedicated, seamless and visually intuitive platform, let alone ground such a mixed-methods approach in a methodological framework sufficient to epistemologically stich them together. Hence the purpose of COMPLEX-IT.

**II. Case-Based Complexity**

COMPLEX-IT is grounded in one of the major epistemological/methodological approaches for study complex social data: *case-based complexity* (CBC) [3-5]. Some of the most widely used CBC techniques include cluster analysis, machine intelligence [6], dynamic pattern synthesis [7] and Ragin's qualitative comparative analysis or QCA [8].

Regardless of the method used, CBC is anchored in four core epistemological arguments that deeply resonate with the majority of computational methods used today, as well as most users in the applied and public sectors. First, the case and its trajectory across time/space are the focus of study, not the individual variables or attributes of which it is comprised. Second, cases and their trajectories are treated as composites (profiles), comprised of an interdependent, interconnected sets of variables, factors or attributes. Third, the relationships and social interactions amongst cases are also important, as are the hierarchical social contexts/systems in which these relationships take place. And, finally, cases and their relationships and trajectories are the methodological equivalent of complex systems – that is, they are emergent, self-organizing, nonlinear, dynamic, network-like, etc – and therefore should be studied as such.

The specific CBC approach that COMPLEX-IT employs is called *case-based modeling*. The utility of case-based modeling is that it is a mixed-methods, computationally grounded approach to learning about and exploring complex social topics and datasets, including big data [9-11]. The methodological platform for case-based modeling is the *SACS toolkit* (sociology and complexity science toolkit), which provides a series of methodological steps and techniques (as well as a mathematical justification) for modeling complex systems in case-based terms [9]. Also, in line with CBC, the purpose of the SACS Toolkit is to model multiple trajectories (particularly across time/space) in the form of major and minor trends; which it then visually and statistically data mines for both key global-temporal dynamics and unique network-based relationships. The SACS Toolkit also data mines its results to either forecast or predict novel cases or trends, as well as simulate different case-based scenarios. (For an in-depth overview of the SACS Toolkit, including its mathematical foundation, see https://www.art-sciencefactory.com/cases.html).

**III. COMPLEX-IT: A case-based approach to social inquiry**

Given the above methodological challenges, the purpose of COMPLEX-IT is to employ the case-based methodological framework of the SACS Toolkit to make the otherwise complex tools and techniques of computational modeling accessible to a wider audience who may have less experience with them. To do that, COMPLEX-IT improves the user-centeredness of data mining by distilling these techniques into their essential features and streamlining their integration– which is accomplished in two key ways: functionality and interface design. COMPLEX-IT's *functionality* is unique because it runs a specific suite of techniques that support case-based data exploration, modeling, forecasting and scenario simulation. In turn, COMPLEX-IT's *tab-driven interface* provides users a seamless, concise and visually intuitive platform. Also, advanced users can examine, download or modify COMPLEX-IT's algorithms, results, and code.

Currently (circa 2019), COMPLEX-IT's suite includes (1) k-means cluster analysis, (2) the Kohonen topographical neural net, (3) a series of data visualization techniques, (4) a machine intelligence algorithm for data forecasting, and (5) a tab for simulating and exploring future scenarios. The fifth tab, in particular – *case-based scenario simulation* (CBSS) – is a major advance in the methodological literature, as it provides an alternative to agent-based modeling and microsimulation for exploring how to influence, change or alter the course of complex data/systems.

*A. Case-Based Scenario Simulation* (CBSS)

The purpose of CBSS is to create a simulated environment for users to visually and statistically explore different possible scenarios and outcomes for some set of case-based clusters/trends, which have been identified earlier in the data analysis process. To do so, CBSS draws inspiration from and builds upon three methodological traditions: (1) microsimulation and agent-based modeling; (2) case-based complexity and case-based modeling; and (3) scenario analysis and planning.

First, in terms of microsimulation and agent-based modeling [12-13], CBSS models multiple cases and their evolving trajectories across time/space. The difference, however, is that CBSS specifically focuses on the mesoscopic level, examining case-based clusters and trends whereas microsimulation and agent-based modeling focus on microscopic un-clustered cases.

In terms of case-based complexity [9-11], CBSS leverages k-means cluster analysis [14 as a user-driven way to identify major and minor clusters/trends among a set of cases. The case cluster/trends identified by k-means are then corroborated and extended through the self-organizing map (SOM), an artificial neural network technique that preserves the topography of analyzed data and which is commonly used in conjunction with k-means [9, 15]. From here, results are explored using a series of data visualization tools, in particular the SOM output grid (See Fig 4 below).

Finally, we draw on scenario analysis and planning [16-17], a broad collection of techniques for developing and evaluating scenarios effects on an entity of interest. Evaluating these scenarios provides insight into how the entity might respond under uncertain circumstances and informs planning [18]. In CBSS, the 'scenario simulation' component enables targeted exploration of how case-based clusters/trends respond to various plausible scenarios they may encounter. These *scenarios* can range from strategic interventions in the complex data/systems of study to external events affecting them. Thus, summarizing from above, CBSS offers users several strengths in the study of complex data/system.

- It allows users to explore how a cluster/trend can be driven in a more desirable direction, by simulating different scenarios or interventions into its composite of causal conditions (i.e., its profile of factors, variables, measurements) and then running the model to see if the desired change took place. In other words, given its emphasis on learning, it can generate multiple and different models of complex data/systems that are flexible and evolving to the needs of the user. Generating multiple models is useful when there (1) is a high level of uncertainty around a cluster/trend of interest; (2) there are multiple plausible interventions that can take place; or (3) there are multiple events that could impactfully affect the cluster/trend.

- In addition to exploring what leads to a desired change, scenario simulations help the user learn about (1) how different clusters/trends respond to plausible events; (2) how resilient they may be to changing classification in the state space; (3) what leads to undesirable change; and (4) the type and degree of intervention necessary to propel a trend toward another cluster/trend's profile.

- It offers the ability to analyze how different clusters or the entire complex system of study might react to various possible scenario changes or interventions in order to help users plan for the multiple contingencies and paths the cluster/trends and system face.

- Finally, unlike agent-based modeling, CBSS always empirically dependent and driven, starting with the user's data (Castellani, Barbrook-Johnson & Schimpf 2019). In other words, one has to use data to employ the CBSS approach.

Despite these advantages for the study of complex data/systems, CBSS has some important limitations. First, being empirically driven means CBSS analysis cannot depart far from the data it uses, unlike other simulation techniques (e.g., agent-based modeling). Also, CBSS does not model the interactions amongst a set of cases or their corresponding clusters or trends. However, we are presently developing a case-based ABM tab [19]. Finally, CBSS cannot easily be used for forecasts or scenarios into the distant future, as there are too many contingencies and uncertainty to know where systems may be after a substantial gap of time.

**Implementation and architecture**

*Fig. 1 COMPLEX-IT Interface*

COMPLEX-IT (shown in Fig 1) was built with the R programming language and Shiny, a web-framework for R. It can be accessed through the web for the server hosted version or downloaded and run with a localhost through the RStudio IDE. We use a Unified Modeling Language (UML) activity diagram to depict its architecture [20]. Activity diagrams present a series of activities a system progresses through over a use-session, including branching paths different sessions may take.

The activity diagram for COMPLEX-IT is displayed in Fig 2. Note the diagram displays the three core paths we've developed for COMPLEX-IT – the design of which was informed from feedback and requests from different users (particularly those in evaluation research). However, other activity flows can be employed. Within COMPLEX-IT each of the activities in Fig 2 has a specific set of inputs, outputs and designated tasks. Each activity or tab is also accessible as a separate self-enclosed section of the app. Below we present each tab following the order of Fig 2.

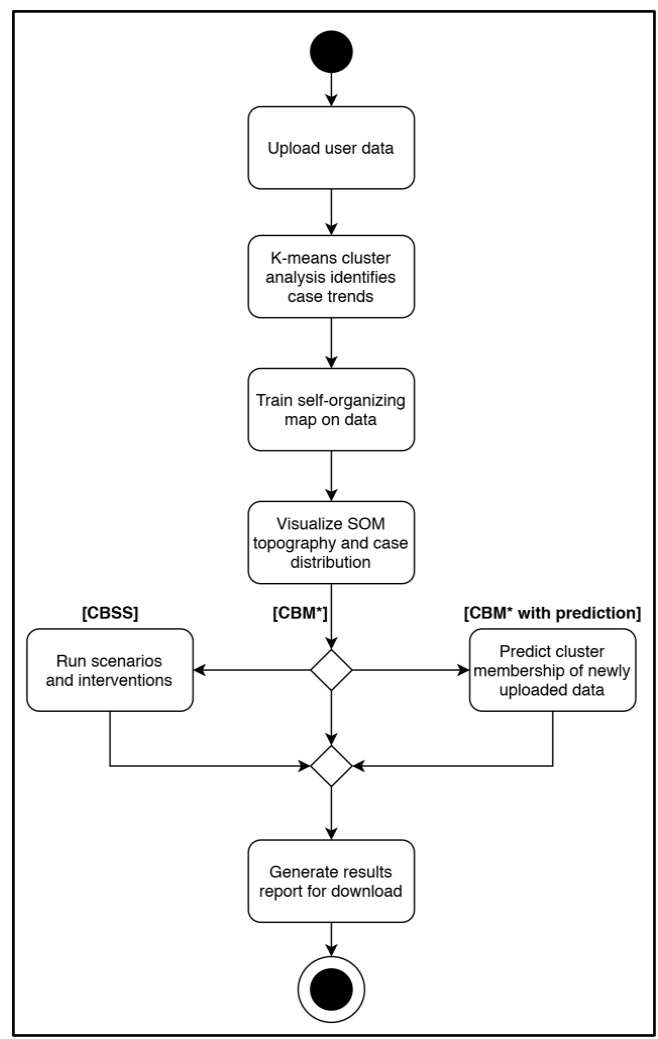

Fig. 2 – UML Activity Diagram for COMPLEX-IT. *Case-based modeling

**I. Data upload tab**

The main purpose of this tab is to enable users to upload and shape their data for analysis. Uploaded data should be in *CSV* format and follow the structure of rows as cases and columns as the profile components of the cases. Only data with numeric values can be analyzed in COMPLEX-IT. Upon uploading data, an adjustable preview of the rows and columns will appear at the bottom of the dashboard. Finally, users can modify which elements are included in cases' profiles before starting analysis by sub-setting the uploaded case data.

**II. Cluster your cases tab**

The main purpose of this tab is to enable users to group cases together in order to identify major and minor case clusters/trends. This tab employs the k-means clustering algorithm, as a first step toward learning and modeling one's data, because it requires users to apply their experience and knowledge of their topic of study to select an appropriate number of groups or clusters and to evaluate the validity and empirical sensibility of the output (See Fig 3).

*A. Overview of K-means and COMPLEX-IT output*

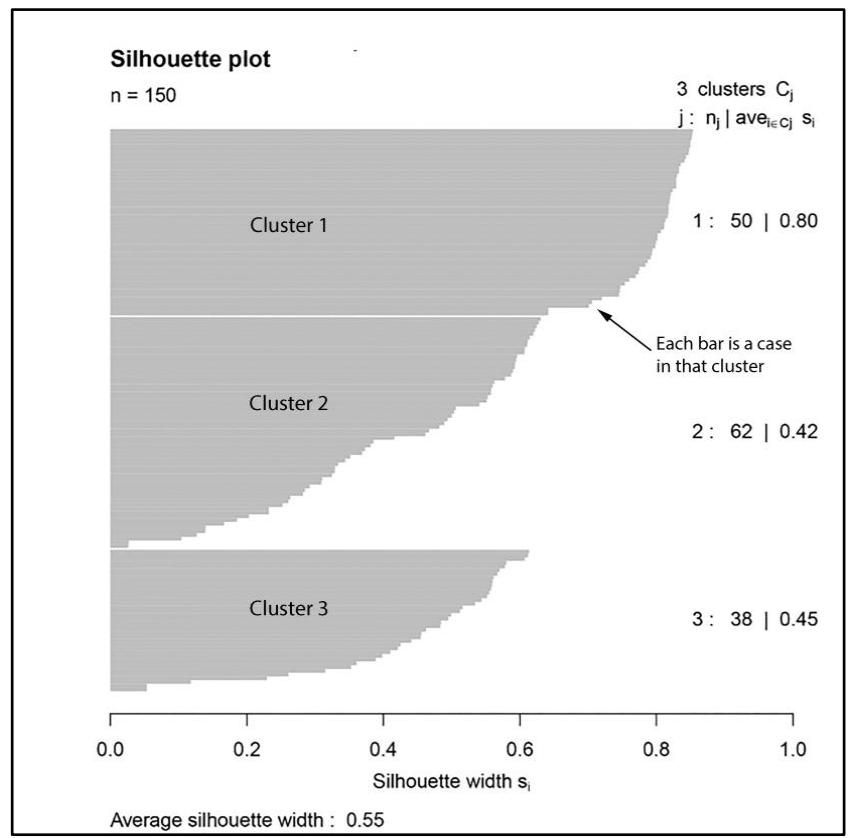

*Fig. 3 – Silhouette plot. Each cluster is plotted as a horizontal bar-plot, with a bar for each case. Bar values span the range of -1 to 1, with near 1 being a strong fit and zero or less being a very poor fit. The 'average width' or fit within each cluster is displayed on the right after the cluster number and size; and the 'overall average fit' is found at the bottom.*

K-means operates by separating cases into groups by minimizing the within group differences between a cluster's associated cases [12]. K-means then iteratively moves cases between groups to minimize within group differences. A final cluster is represented by a centroid, which contains the average values for each element of the case profiles in each cluster. Thus, distributions across clusters ideally are tightly packed around the centroid and unique from other clusters in the set.

Upon running the k-means algorithm, COMPLEX-IT will present the resulting cluster profiles and size of the clusters. Additionally, the user can request evaluations on the quality of the clustering: pseudo f and silhouette plots, two commonly used metrics for evaluating clusters. Pseudo f is an overall measure of how tightly cases are grouped within clusters and how separate or non-overlapping clusters are, with higher values indicating better performance. The silhouette plot allows for visual and quantitative inspection of cases fit within their clusters, see Fig. 3. Through inspection of the clusters vis-à-vis domain knowledge or relevant theory and these quality measures, users can identify the best arrangement for the major and minor trends in their set of cases.

**III. SOM training tab**

The main purpose of this tab and the following tab is to enable the user to employ the self-organizing map (SOM) – an unsupervised neural network clustering algorithm – to corroborate or raise questions about the case clusters/trends identified using k-means. The solution produced by the SOM in Tab 3 (which is stored as an object during a session) is subsequently used to explore different case-based social scenarios (Tab 5) or to classify (predict/forecast) new cases (Tab 6).

*A. Overview of the SOM and its output*

The SOM is a topographical neural network designed for unsupervised learning (i.e., machine intelligence), data visualization, and clustering [21]. This value of this approach is that it puts the burden of identifying the major and minor trends on the algorithm, in contrast to user-driven k-means. Therefore, the SOM solution can be compared to the k-means solution to corroborate results or suggest further analysis.

We chose SOM over other neural network techniques because it is a well-established and popular approach with unique features for supporting human analysis of results. Unlike other artificial neural network techniques, the SOM classification model can be directly visualized and analyzed to understand how cases were assigned to a neuron (cluster) outcome – which is the focus of tab 4. The classification model is directly interpretable because the SOM projects high-dimensional (multiple variable) input data onto a 2-dimensional grid solution (See Fig 4 below) that can be visualized in a variety of ways – which we discuss in the SOM analysis section.  Thus, the SOM provides users, including those less familiar with machine learning techniques, a more direct way to compare their k-means and SOM results.

*B. Options on SOM Tab*

The options on this tab concern running the SOM algorithm, including setting the size of the grid, weights initialization, the number of algorithm iterations, data scaling, and a seed for preserving the initialization and the learning rate. After training the SOM, ANOVA is run to provide information on which (if any) of the profile elements differed significantly across the neurons. This lends insight into what may be distinguishing factors across the neurons. Additionally, two quality measures are displayed, the *quantization error* and the *topographical error*. The first indicates how much, on average, cases diverge from their assigned neuron; and the second addresses the rate at which surrounding neurons are a good fit for a neuron cases across the grid. Preferably these have lower values close to zero, as they are measures of error.

**IV. SOM analysis tab**

The main purpose of this tab is to enable the user to corroborate the case trends identified in the SOM neuron profiles with the k-means cluster profiles and to visually explore the results of their combined methods of analysis.

*A. Understanding the Tab 4 SOM Map*

The central object of study in Tab 4 is the SOM map, shown in Figures 4 and 5. As recalled from earlier, the SOM 'maps' complex data by creating a grid of *n* x *n* neurons. On this map, each neuron has a set of weights equal to the number of configurational variables in the case profile (i.e., causal conditions, factors, measures, etc.), which makes each neuron somewhat similar to a k-means cluster centroid. For example, looking at Figures 4 and 5, one sees a *5 x 5* SOM Map, with a total of 25 neurons (i.e., cluster solutions). We happen to call these neurons *quadrants*, which came from our work with various users, who repeatedly used the word.

One also sees in Fig 4 that some of the quadrants do not have a bar-plot configuration. The same with Fig 5: some of the quadrants do not have cases. The reason is that these quadrants do not constitute viable cluster solutions for the data. And it is because of these 'empty' quadrants, in part, that the SOM is referred to as *unsupervised learning*, as it lets a dataset of cases settle (map) onto the 'best' cluster solution across all quadrants, with cases iteratively finding the 'right' quadrant on which to live. Unsupervised learning takes place by the SOM setting default weights for all the quadrants and then associating cases with the quadrants they most closely resemble, usually through a distance measure. The weights of the quadrants and their neighbors are then updated based on an adjustable learning factor, which also leads to similar cases 'settling' into similar regions of the map. The degree of 'learning' for each quadrant (and its capacity to attract similar cases in its region) decreases over time and the SOM is assessed, as mentioned earlier, by the two validity measures *quantization error* and the *topographical error*.

In short, the SOM's 'topographical' mapping of the data leads to quadrants in proximity having similar weights for their configuration of variables (i.e., factors, causal conditions, measures, etc.) and those further away having progressively larger differences in weights. In other words, quadrants (and hence cases) that are near each other on the map are more alike configurationally than those further apart. Hence why this approach is called the self-organizing topographical map.

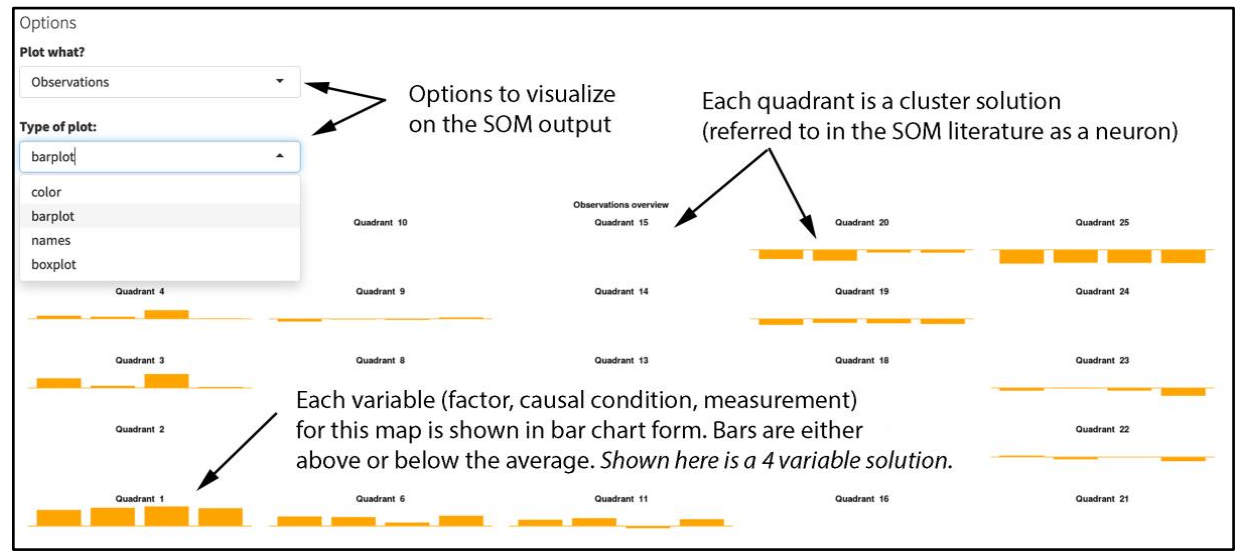

*Fig. 4 SOM Bar-plot. Bars represent the average value for case elements in each quadrant.*

*B. Options on SOM Map Tab*

As suggested above, the strength of tab 5 is the variety of options it provides for visualizing and analyzing the SOM's results. They are broken down into two types: prototypes (i.e., the configurational factors/variables in one's study) and observations (i.e., cases in the dataset).

Fig 4, for example, is a bar-plot of a study exploring four factors. The bar-plot displays the entire SOM grid and each quadrant will contain a series of bars or no bars if no cases were assigned to that particular quadrant. Each bar in a quadrant represents one of the elements of the cases' profiles and should be read from left to right. Similar to k-means, the bars for each quadrant represent the average values for cases within it. Bars are centered on the global mean for any given case variable; therefore, if a bar is only a line within the quadrant, it is at the global mean, whereas bars higher or lower than the center line indicate it is higher or lower than the global mean, respectively. In this way the bar-plot provides a view into the major and minor trends for the cases through quadrant profiles. As some quadrants may have similar profiles, full numeric data for each of quadrant can be retrieved from the Generate Report tab. Quadrant profile data can then be used to group quadrants based on similarity and thereby simplify the set of major and minor trends.

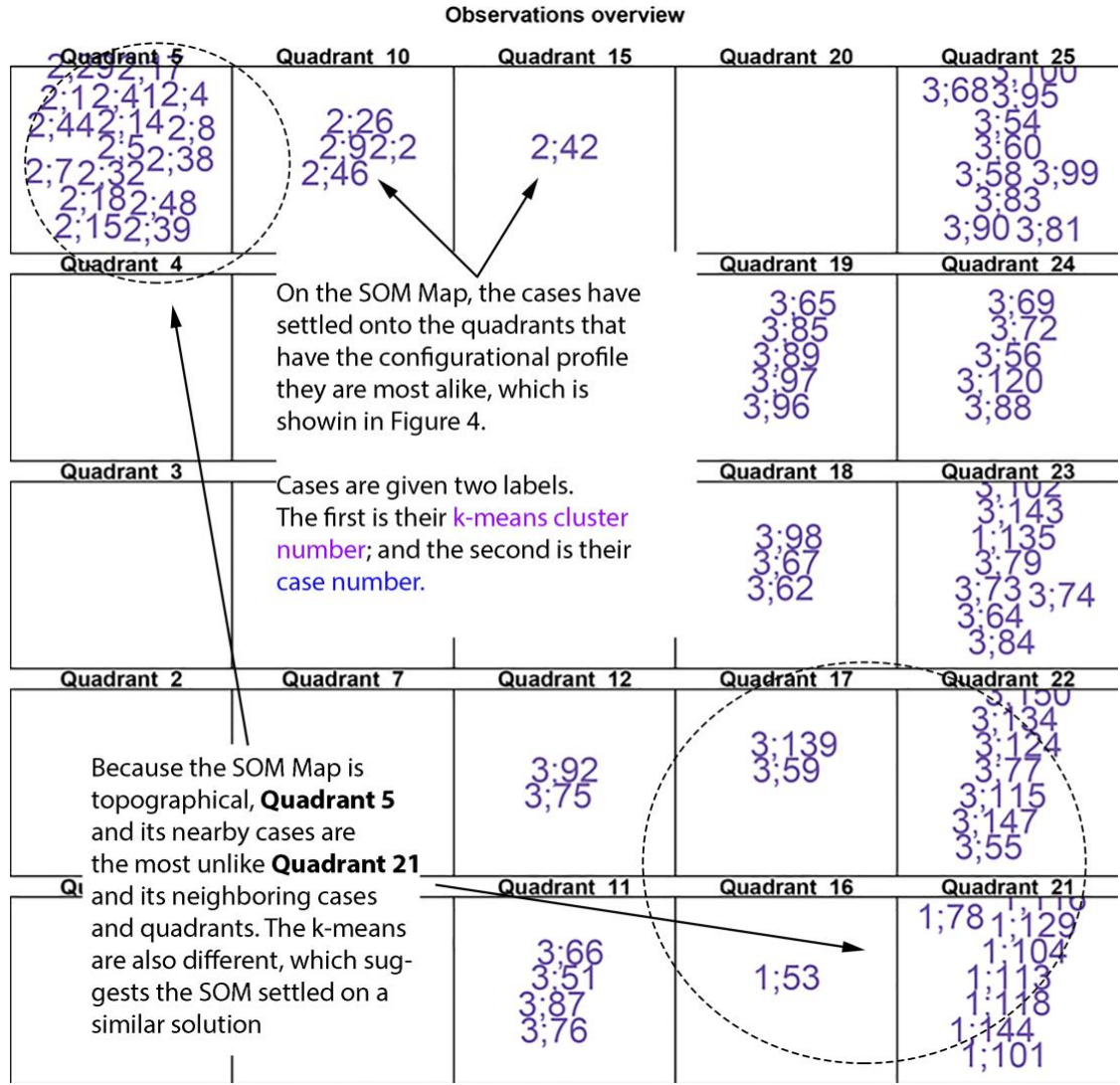

Fig. 5 SOM Names plot. Note, for each plotted case cluster ID is first and case ID is second

C. Corroborating the SOM and K-means Solutions

Corroboration begins by relating the major and minor trends identified in the SOM quadrant profiles back to the k-means cluster profiles. The global mean values for each case element are also listed in the k-means cluster table as a reference. This process may also raise questions and a need to revisit one or both of the clustering approaches if there are wide discrepancies between them. The names plot can then be used, which displays the k-means cluster ID and case ID for each case on top of the SOM grid. Users should be aiming for good convergence between the k-means cluster and SOM quadrant cases are assigned to; if many cases diverge new clusters may be necessary or the data may be incorrectly specified. Note that if many cases are being analyzed, it may not be possible to print of the cluster and case IDs on the grid. Full results can be retrieved from the Generate Report tab, however.

**V. Simulating scenario/intervention tab**

Simulation is a powerful exploration tool, particularly for users in the public sector, policy evaluation, or applied fields of study, where having a risk-free environment to understand how, when, and where effective interventions can be made for some population or complex system of study is paramount. Simulation, however, comes with its own challenges. One challenge is staying close to the data of study; another is the need to remain focused on the population of cases (as opposed to variables) in a study; and, finally, there is the need to build a model that is simple and yet complex enough to extract meaningful information. Hence the purpose of case-based scenario simulation (CBSS). Given we reviewed the purpose and aims of CBSS in the introduction (Section 3A), we will focus here, instead, on how Tab 5 works.

*A. The goal of CBSS*

The main purpose of Tab 5 is to enable users to simulate different scenarios that the major and minor clusters/trends in their dataset could be driven to experience, as well as any associated contingencies or responses and dynamics or failures to change. Here we define 'interventions' as a particular type of simulated (but potentially real) scenario reflecting some proactive attempt to change one or more of the configurational factors for a cluster or trend (i.e., variables, measurements, causal conditions). Examples of interventions could be wider changes in the settings or systems in which the clusters are situated (i.e., economic or environmental changes); or they could be specific policy changes (i.e., improve schooling or access to health care) or changes within the cluster by its own cases (i.e., a community-level shift in thinking or behavior). Other scenarios a user may wish to explore include reactive events originating from environmental forces or outside of proactive action, such as out-migration in a community. Better understanding from scenario simulations can also lead to more informed planning or strategic action, which may be desirable for areas like policy analysis and program evaluation.

*B. Understanding the Tab 5 SOM Map*

To begin, CBSS combines results from the previously trained SOM and k-means analysis, presenting the same SOM map used in Tab 4. However, unlike the SOM Map in Tab 4, which plots all the cases in the study, Tab 5 plots the k-means clusters initially identified in Tab 2.

CBSS takes this focus because the k-means solution offers a more concise user-driven summary. For example, if one had a study with N=100 cases, simulating interventions into each and every one of them in Tab 5 may prove unmanageably complex and time consuming. Still, given that the SOM map in Tab 4 exists, the cases for any given cluster and their respective differences can always be examined. In other words, while exploring the CBSS map in Tab 5, one can also explore the SOM

Map in Tab 4 to gain a more granular understanding of how interventions or changes made on the CBSS map might differentially 'play out' at the case level.

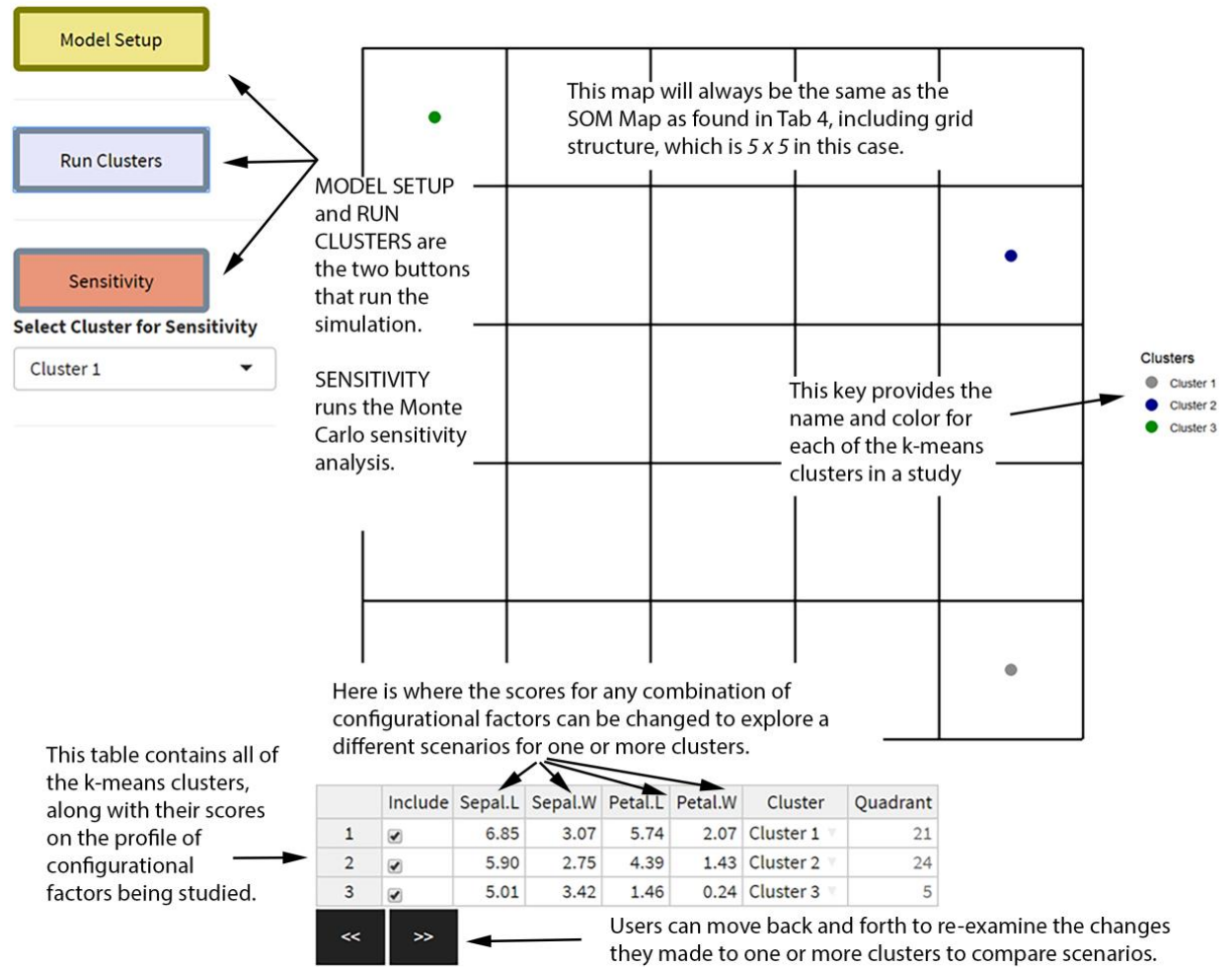

Fig 6 Scenario and Intervention tab main display and controls

*C. Running a simulation*

By pressing the *Model Setup* and *Run Clusters* buttons, the Scenario tab will be initialized, as shown in Fig. 6. Below the SOM plot an editable table allows the user to update one or more elements of the k-means case trends. Different scenarios can be explored by changing the relevant elements in the table for a given scenario and pressing *Run Clusters*. The updated case trend will be re-examined against the SOM quadrants and the impact of the scenario, if any, will be shown through which quadrant it is now most closely associated with. The SOM bar-plot is also plotted here as a reference for the quadrant profiles.

Exact estimation of the change a scenario causes a case trend may not be possible. Therefore, the Scenario tab also offers sensitivity analysis to account for uncertainty in the efficacy of scenario changing a case trend. The user can specify how much a change to one or more element in an updated case trend may deviate by, from zero to one hundred percent of change compared to the original values of the case trend, both in the positive and negative direction. This can be accessed by selecting the case trend or cluster to test and pressing the *Sensitivity* button on the left panel. After inputting the deviation for changed element(s), a Monte Carlo simulation will randomly sample from across this range, testing and providing a summary of which quadrants the case trend would be associated with – which allows a key insight into how 'sensitive' the effect of the scenario is to projected deviations.

### VI. Prediction and classification tab

Tab 6 was created to help users involved in fields where they are regularly asked to make predictions or forecasts regarding 'new' or 'different' cases not found in their complex datasets or, alternatively, the same cases at a different point in time (as regularly explored in longitudinal or time-series data). Examples range from identifying 'at risk' urban communities due to air pollution exposure or decreasing access to healthcare to predicting what sorts of social services a new client or patient might need.

Like Tab 4 and Tab 5, this tab uses the trained SOM as its map. However, the focus here is back to the cases and the SOM map quadrants. In other words, the goal is to decide which map quadrant to which some new case or set of cases belongs, based on the SOM solution arrived at in the initial dataset studied.

The process is rather straightforward. Similar to the Data Upload tab, data can be uploaded to this tab. It is critical, however, that the uploaded data used in this tab have the same data columns as the data originally uploaded in COMPLEX-IT for any given session. Otherwise COMPLEX-IT will reject the new data and display a warning, as the SOM cannot make predictions on variables that were not part of its training set. After classifying the new dataset, a table will display the cases as well as their best-fitting and second best-fitting quadrant; the bar-plot from the SOM is also displayed as a reference for the quadrant profiles.

### VII. Generate report tab

The main purpose of this tab is to generate an adaptive set of exportable results. The generate report tab can be accessed any time to export detailed results from the k-means and SOM analysis as well as the scenario and predict tab. Results will only be exported for those tabs the user has employed in the current session; so long as at least one tab has been used it is possible to export results. Exported results will reflect the most recent analysis, simulated scenario, or prediction performed.

More comprehensively, k-means results include the cluster profiles, pseudo f, cluster ID for all cases and the silhouette. SOM results include the parameters of the SOM (e.g., learning rate), the ANOVA results and quality measures, the quadrant profiles, quadrant IDs for cases as well as the bar-plot and boxplot graph. Scenario results include the intervention tested and sensitivity analysis results while the prediction results include the new data and their assigned SOM quadrant. Exportable results are intended to allow the user to share or disseminate findings or conduct further analysis.

**Quality control**
COMPLEX-IT has been developed through an agile process over the past three years, with team members regularly meeting, planning, and implementing new features which were then reviewed and revised as the platform evolved. During this time, the development team tested new features as they were introduced to ensure they performed as expected.  These were subject to regular full platform reviews to check that all components were properly integrated. Additionally, the primary team members regularly analysed new datasets through COMPLEX-IT. Any technical bugs, interface difficulties or other challenges were reported to the development team and added to the ongoing meetings to resolve.

**(2) Availability**

*Operating system*
The hosted version of COMPLEX-IT accessed through a browser and is compatible with most modern web browsers. The downable version of COMPLEX-IT is compatible with Windows 7 or MacOS 10.7 or later versions of Windows or Mac. It is also available on Debian, SUSE, Ubuntu and Redhat Linux distributions.

*Programming language*
R 3.3 or above.

*Additional system requirements*
No additional requirements for the hosted version of COMPLEX-IT. If users want to run locally, it will be necessary to installed R and RStudio, which require a minimum of 256 megabytes of RAM.

*Dependencies*
All dependencies are packages or frameworks for R.
cluster 2.0.7.1 or higher
ggplot2 3.1.0 or higher
rhandsontable 0.3.7 or higher
shiny 1.2.0 or higher
shinyThemes 1.1.2 or higher
SOMbrero 1.2.3 or higher
zip 1.0.0 or higher

*List of contributors*
Brian Castellani project lead, researcher, and developer
Corey Schimpf lead developer and researcher

Mike Ball server administrator and developer
Peter Barbrook-Johnson project mentor and researcher
Nigel Gilbert project mentor

***Software location:***
    **Code repository**
        ***Name:*** GitHub
        ***Identifier:*** https://github.com/Cschimpf/Complex-It
        ***Licence:*** MIT
        ***Date published:*** 17/05/19

***Language***
English

### (3) Reuse potential

Case based modeling and scenario simulations hold great potential for social scientists of all backgrounds interested or already involved in the study of complex data/systems. Furthermore, policy analysts and program evaluators may benefit from the platform, for instance by analyzing how a set of systems are distributed before and after a real-world policy or program intervention and/or analyzing how these systems might respond to external events or new interventions through scenario simulation. COMPLEX-IT may also be used as an exploratory, datamining tool to identify variations in complex data/systems, trajectories and responses to change.

**Acknowledgements**
The authors wish to thank Centre for the Evaluation of Complexity Across the Nexus (CECAN) for their financial and intellectual support, as well as Durham University for its financial support. Finally, we would like to thank Carl Dister for his development support in the early stages of COMPLEX-IT.

**Funding statement**
COMPLEX-IT was developed with support from the grants received by CECAN, namely, Economic and Social Council (grant numbers: ES/N012550/1 and ES/S000402/1). Funding was also received from Durham University via the ESRC pathways to impact grant.

**Competing interests**
The authors have no competing interests to declare.